\documentclass[times]{aa}
\usepackage{psfig}
\usepackage{natbib}
\usepackage{epsf}
\usepackage{graphicx}

\bibpunct{(}{)}{;}{a}{}{,}

\begin{document}

\def\eps{\varepsilon}
\def\aap{A\&A}
\def\apj{ApJ}
\def\apjl{ApJL}
\def\mnras{MNRAS}
\def\aj{AJ}
\def\nat{Nature}
\def\aaps{A\&A Supp.}
\def\prd{PRD}

\def\me{m_\e}
\def\lesssim{\mathrel{\hbox{\rlap{\hbox{\lower4pt\hbox{$\sim$}}}\hbox{$<$}}}}
\def\gtrsim{\mathrel{\hbox{\rlap{\hbox{\lower4pt\hbox{$\sim$}}}\hbox{$>$}}}}
\def\cmb{{\rm cmb}}
\def\a{\alpha}
\def\e{{\rm e}}
\def\o{{0}}
\def\p{{\rm p}}
\def\me{m_{\rm e}}
\def\gh{{\rm gh}}
\def\cl{{\rm cl}}
\def\rg{{\rm rg}}
\def\th{{\rm th}}
\def\rel{{\rm rel}}
\def\jet{{\rm jet}}
\def\power{{\rm power}}
\def\aC{{a_{\rm C}}}
\def\ab{{a_{\rm b}}}
\def\as{{a_{\rm s}}}
\newcommand{\hq}{\hbar}
\newcommand{\epsB}{\varepsilon_{\rm B}}
\newcommand{\epsCMB}{\varepsilon_{\rm cmb}}

\def\CR{{\rm rel}}
\def\gal{{\rm gal}}
\def\gas{{\rm gas}}
\def\rel{{\rm rel}}

\def\del#1{{}}

\def\C#1{#1}

\title{Searching for the relativistic Sunyaev-Zeldovich effect with
  the PLANCK experiment}
\titlerunning{Relativistic Sunyaev-Zeldovich effect with
  PLANCK}
\author{Torsten A. En{\ss}lin$^1$ \and Steen H. Hansen$^2$}
\authorrunning{En{\ss}lin \& Hansen}
\institute{Max-Planck-Institut f\"{u}r Astrophysik,
Karl-Schwarzschild-Str.1, Postfach 1317, 85741 Garching, 
Germany \and  University of Zurich, Winterthurerstrasse 190,
8057 Zurich, Switzerland} 

\abstract{Populations of relativistic electrons in clusters of galaxies should
  imprint a characteristic spectral distortion signature on the Cosmic
  Microwave Background (CMB) photons -- the relativistic Sunyaev-Zeldovich
  (SZ) effect. We investigate how sensitive PLANCK and successor experiments
  will be in detecting or constraining the total optical depth $\tau_\CR$ of
  relativistic electrons. We find an expected sensitivity of PLANCK of
  $\sigma_{\tau_\CR} \sim 2\cdot 10^{-5}$ for an individual cluster, which is
  too insensitive to detect even optimistic scenarios, which predict $\tau_\CR
  \sim 10^{-6}$ to $10^{-7}$. However, by stacking the PLANCK-signal of all SZ
  detectable clusters the relativistic SZ effect can be statistically probed
  to a sensitivity level of $\sigma_{\tau_\CR} \sim 10^{-7}$. PLANCK successor
  experiments will be able to probe even conservative scenarios of
  relativistic electron populations in clusters of galaxies.
 } \maketitle

\section{Introduction\label{sec:intro}}

Clusters of galaxies contain a several keV hot thermal plasma, which
is able to Comptonize the CMB while its photons are traversing the
cluster. The resulting observable spectral distortions of the CMB --
the SZ effect \citep{1972ComAp...4..173S} -- are discriminated by the
different nature of the energy forms of the scattering electrons: the
thermal energy manifests itself in the thermal SZ (tSZ) effect, and
the kinetic energy of a gas bulk motion in the kinematic SZ (kSZ)
effect. The spectral characteristics of the tSZ and the kSZ effects
are independent of the electrons' energies as long as the electrons'
velocities are much smaller than the speed of light $c$. However, the
thermal electrons in hot galaxy clusters with temperatures of $\approx
10$~keV have velocities of the order of $0.2\,c$, and therefore
relativistic corrections to the canonical tSZ spectral distortions
have to be taken into account~\citep{1979ApJ...232..348W}.

In addition to the mildly relativistic thermal electrons there are
also highly relativistic electrons present in galaxy clusters, which have
their own characteristic relativistic SZ (rSZ) spectral distortion
signature \citep{2000A&A...360..417E}. It is the aim of this work to
investigate how future sensitive spectral measurements of CMB
distortions caused by galaxy clusters can be used in order to
investigate or constrain such electron populations.

We know about the existence of highly relativistic electrons in galaxy
clusters since extended, cluster-wide radio emission, the so called
{\it cluster radio halos}, has been detected in roughly one third of
the X-ray brightest galaxy clusters~\citep{Giovannini.Pune99}. The
radio halo emission is produced by radio synchrotron emitting
ultra-relativistic electrons with energies of the order of
10~GeV. Relativistic electrons at energies far below a GeV have
escaped our measurements so far since their synchrotron emission is at
frequencies below the Earth ionosphere transparency cutoff, although
their dim synchrotron self-Comptonized radiation should fall into our
telescope's frequency bands \citep{2002A&A...383..423E}. However,
large numbers of such relativistic electrons are expected, not only as
a low energy continuation of the observed radio emitting electron
spectra, but also due to numerous physical generation mechanisms. Such
mechanisms are cluster merger shock waves, radio galaxy outflows, and
electron production in inelastic interactions of relativistic protons
with the thermal gas nucleons \citep[][ for
reviews]{Pune99,astro-ph/0208074}.  There are two very different
habitats of relativistic electrons in galaxy clusters. The first is
the relatively dense thermal gas, in which Coulomb cooling should
quickly thermalize any significant electron population below 100 MeV
\citep{1999ApJ...520..529S}.  The second are the fossil remnants of
former radio galaxy cocoons, the radio quiet ghosts cavities
\citep{Ringberg99}, in which even MeV electrons may survive for
cosmological times due to the low gas density there.  The physical
origin of any high energy electrons is very interesting, and since our
present observational knowledge about these MeV to GeV electrons is
poor \citep[][ for a compilation of the available spectral
information on Coma]{1998AA...330...90E}, any means to obtain additional
information is highly desirable. In this paper we will analyze the
ability of the PLANCK experiment or its successors to detect or
constrain such electrons.

In contrast to a thermal electron population%
%
%
\footnote{The reported high energy X-ray excess of the Coma galaxy
  cluster \citep{1999ApJ...513L..21F} triggered some speculations
  about the existence of a supra-thermal electron pupulation in
  between 10 and 100~keV, which also would produce a unique SZ
  signature \citep{1999AA...344..409E, 2000A&A...360..417E,
  2000ApJ...535L..71B, 2000ApJ...532L...9B, 2002MNRAS.337..567L,
  2003A&A...397...27C}. However, such a population is questioned on
  theoretical grounds \citep{2001ApJ...557..560P}, and even the high
  energy X-ray excess itself is under debate
  \citep{astro-ph/0312447,astro-ph/0312625}.}%
%
%
, the spectral shape of a relativistic electron population is poorly
known. At GeV energies, they are well described by power
laws. However their number density is generally dominated by the
unknown lower end of their spectral distribution, which is expected to
have a broad peak. Since the rSZ effect depends practically only on
the total number of relativistic electrons and very little on their
spectral distribution we approximate the spectrum by a delta function
centered on that spectral peak.  As will become clearer in the
following, this idealized treatment is sufficient for our
investigations of the sensitivity to detect the rSZ effect. A more
realistic treatment would introduce several poorly constrained
parameters like spectral index and low energy cutoff and its detailed
shape, for which the natural choice is not obvious.

\section{Spectra of CMB distortions\label{sec:spectra}}

The PLANCK experiment should be able to identify $\sim 4\cdot 10^4$ clusters of
galaxies via their tSZ effect \citep[e.g.][]{2001A&A...370..754B}. In order to
be able to do spectral precision measurements of the CMB distortions caused by
a galaxy cluster one has to suppress other contaminations as far as
possible. A way to do this is optimal spatial filtering, which allows to
remove efficiently foregrounds like galactic dust and synchrotron emission,
but also CMB temperature fluctuations, which have their spatial power mostly
on large angular scales \citep{1996MNRAS.279..545H}. In the following we
assume that such a filtering step was applied to the frequency maps of PLANCK,
leading to galaxy cluster spectra which are mostly free of such
contaminations. Since the galactic synchrotron and dust emission might have
some power on small angular scales, which would partly survive such a
filtering step, we further assume that the lowest and highest frequency
channels of PLANCK are used to remove these residual contaminations.

The CMB-blackbody spectrum 
\begin{equation}
I_\nu = i_0\, i(x) = i_0 \; \frac{x^3}{e^x -1}\, ,
\end{equation}
with $x = h \nu/kT$ and $i_0 = 2\,(kT_\cmb)^3/(hc)^2$ becomes
distorted due to the inverse Compton scattering off the electrons in a
galaxy cluster. The spectral distortions are described by
\begin{eqnarray}
\delta i(x) &=& g(x) \,y_\gas \, \left( 1+\delta(x,T_e) \right) 
- h(x) \, \bar{\beta}_\gas \tau_\gas  \nonumber \\
&&+ (j(x,p)- i(x)) \, \tau_\rel \,.
\label{eq:deltai(x)}
\end{eqnarray}
The first rhs term describes the tSZ distortions, which have a shape given by
\begin{equation}
g(x) = \frac{x^4 \,e^x}{(e^x -1)^2} \left( x\,\frac{e^x + 1}{e^x -1} -
4 \right)\,,
\end{equation}
and a magnitude described by the the Comptonization parameter 
$y_\gas = {\sigma_{\rm T}}/({\me\, c^2})\, \int\!\! dl\, n_{\e ,\gas}
\,kT_\e$ .
For non-relativistic electrons the relativistic correction term is
zero, $\delta(x,T_e) = 0$, but for hot clusters even the thermal
electrons are slightly relativistic, which will modify the thermal SZ
effect~\citep{1979ApJ...232..348W}.  These corrections are easily
calculated~\citep[see e.g.][]{1995ApJ...445...33R,
2003astro.ph..7519I, 2000A&A...360..417E, 2001ApJ...554...74D}, and
can be used to measure the cluster temperature purely from SZ
observations~\citep{2002ApJ...573L..69H}. For the PLANCK channels this
is most easily implemented using the simple fitting formulae presented
in \citet{2003MNRAS.338..796D}.

The second rhs term in Eq. \ref{eq:deltai(x)} gives the kSZ
distortions, which have the spectral shape
\begin{equation}
h(x) = \frac{x^4 \,e^x}{(e^x -1)^2}
\end{equation}
and depend on $\bar{\beta}_\gas$, the average line-of-sight streaming
velocity of the thermal gas ($v_\gas = \beta_\gas \, c $, $\beta_\gas
>0 $ if gas is approaching the observer), and the Thomson optical
depth
$\tau_\gas = \sigma_{\rm T} \int\!\! dl\, n_{\e ,\gas} .$

Finally, the last rhs term in Eq. \ref{eq:deltai(x)} describes the
distortion by relativistic electrons, which we approximate to be
mono-energetically distributed.  We characterize the electron
energy via the dimensionless momentum $p = P_\e/(m_\e\,c)$, where
$P_\e$ is the usual electron momentum. The optical depth
$\tau_{\rel} = \sigma_{\rm T} \int\!\! dl\, n_{\e, \rel}$
determines the fraction of strongly scattered photons. Here, $n_{\e,
\rel}$ is the number density of relativistic electrons. The spectral
distortions $\delta i_\rel= (j(x,p)-i(x))\tau_\rel$ are described by
two terms: The IC scattering redistributes the CMB photons, which
means that photons disappear from the CMB spectrum according to
$-\tau_\rel\, i(x)$, but reappear at different frequencies like
$\tau_\rel\, j(p,x)$, which depends on the electron momentum. In
general, $\delta i_\rel$ has to be calculated numerically
\citep{2000A&A...360..417E}, however the non- and ultra-relativistic
limits can be treated analytically.

In the non-relativistic limit ($p\ll 1$), a mono-energetic electron
population leads to spectral distortions which are identical to those
of the well-known tSZ effect with
\begin{equation}
  \delta i_\rel = \frac{p^2}{3}\, g(x)\,\tau_\rel.
\end{equation}
In the ultra-relativistic limit ($p\gg 1$) CMB photons simply
disappear from the spectral range of interest since they are
scattered to very high frequencies, leading to
\begin{equation}
  \delta i_\rel = - i(x)\,\tau_\rel.
\end{equation}

As long as all electrons are in one of these two limiting cases, the
relative shape of the spectral distortion is completely independent of
the details of the electron spectrum, only the distortion strength is
either proportional to the electrons total energy content
(non-relativistic limit) or their number density (ultra-relativistic
limit).

Thus, only in the trans-relativistic regime ($p\sim 1$) the shape of
the spectral distortions depends on details of the electron
spectrum. Since we are interested in the ultra-relativistic case, our
approximation of mono-energetic electron spectra is fine. By keeping
the full rSZ distortion term $\delta i_\rel= (j(x,p)-i(x))\tau_\rel$
\citep[for which we use the formulae given in][]{2000A&A...360..417E},
we make sure that our formalism tells us when we enter the
trans-relativistic regime. For small $p$ the rSZ spectral distortion
approaches the one of the tSZ effect and thereby our sensitivity to
discriminate relativistic electrons decreases.

An order of magnitude estimate of the expected rSZ effect is in
order. Any scattering of a CMB photon with an ultra-relativistic
electron removes the photon from the typical CMB energy range. The
relativistic electron optical depth $\tau_\rel$ can be expressed in
terms of the tSZ $y$-parameter, if a characteristic ratio of the
relativistic electron to thermal gas energy density $X_{\rel,\e}$ can
be assumed to hold approximately on macroscopic scales throughout the
cluster:
\begin{equation}
  \tau_\rel = \frac{3\,X_{\rel,\e}\,y_\gas}{\gamma_\e-1}\, ,  
\end{equation}
where $\gamma_\e=E_\e/(m_\e\,c^2) = \sqrt{1+p^2}$. Inserting some
optimistic values for a Coma-like galaxy cluster like $y_\gas \sim
10^{-4}$, $X_{\rel,e} \sim 0.03$, and $\gamma_\e \sim 10$ for a
relativistic electron population residing in radio quiet ghost
cavities ($X_{\rel,e} \sim 0.1$, and $\gamma_\e \sim 300$ for
relativistic electrons mixed with the thermal gas), one gets $\delta
I_\rel/I = - \tau_\rel \sim - 10^{-6}$ (or $\delta I_\rel/I \sim -
10^{-7}$ respectively).

\del{
\noindent
{\bf ==========================}\\
{\bf ==========================}\\
{\bf ==========================}\\
{\bf ==========================}\\
}


\section{Extracting the rSZ effect}

Our goal is to extract the rSZ spectral signature from PLANCK
experiment multi-frequency measurements. In order to estimate the
significance of our measurement, we apply a $\chi^2$-statistics and
search for the $\chi^2$-minimum with respect to the unknown
parameters. Minimization of the parameters within the ranges given in
table \ref{tab:1} is done through an extention of the publicly
available code {\tt sasz}~\citep{astro-ph/0310149}, based on the
technique of simulated annealing.  In our case the vector in parameter
space, $x$, is a 5 dimensional vector, $\vec{x} = (y, T_e, v_\gas,
\tau_{\CR},p)$.  We choose a simple simulated annealing exponential
cooling scheme, $T_j = c \, T_{j-1}$, where $c\approx0.8-1.0$, and we
use $T_0=1$ and $T_{\rm final} = 10^{-12}$.

\begin{center}
\begin{table}
\begin{tabular}{cc}
\hline 
Parameter & Allowed range \\ \hline
y & $10^{-7}$  to $10^{-2}$  \\
$T_e$ & 0 to 25 keV \\
$v_\gas$ & -5000 to  5000 km/sec \\
$\tau_{\CR}$ & -$10^{-4}$ to $10^{-4}$  \\
p & 0.01 to 1000  \\
\hline 
\end{tabular}
\caption{\label{tab:1} The allowed parameter range.}
\end{table}
\end{center}

\begin{figure}
\begin{center}
\includegraphics[width=0.5\textwidth]{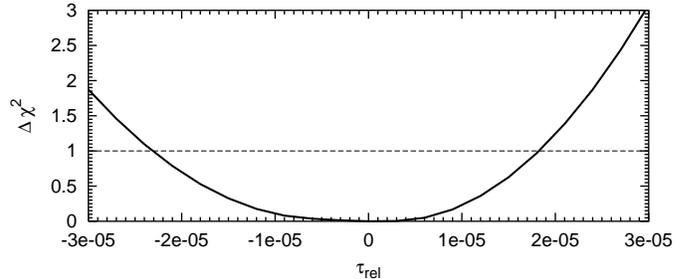}
\end{center}
\caption{$\Delta \chi^2$ as a function of $\tau_\CR$, marginalized 
over the other parameters, $(y,T_e, v_{\rm gas}, p$). $1\sigma$ 
error-bar of $\tau_{\CR}$ is about $2\cdot 10^{-5}$.}
\label{fig:ma}
\end{figure}

In fig. \ref{fig:ma} we plot $\Delta \chi ^2$ as a function of the
unknown $\tau_{\CR}$ for the PLANCK satellite, where we optimistically
assume that only the first and last channels ($30$ and $857$ GHz) are
needed for removal of cluster-scale foreground contamination (like
point sources, and galactic synchrotron and dust emission).  We assume
that the 'true' signal is $\tau_{\CR} = 10^{-6}$ (the other unknown
parameters are $y=10^{-5}, T_e=5$ keV, $v_\gas = 600$ km/sec,
$p=200$), and we maximize over all the unknown parameters.  We find $1
\sigma$ error-bars for the relativistic optical depth $\tau_\CR$ just
about $\sigma_{\tau_\CR} \approx 2\cdot 10^{-5}$.  If one instead excludes
also the $545$ GHz channel, then the error-bars are increased by
approximately a factor of 2.  It is comforting to note that the best
fit point is indeed very near the 'truth' $10^{-6}$.

We next have to test for parameter degeneracy. As is well known, parameter
degeneracy is very important for the error-bars on the normal SZ
parameters. E.g. a large negative peculiar velocity can mimic a higher
temperature (and slightly changed Compton parameter).  In this way a large
negative peculiar velocity of a cluster results in larger error-bars on the
determination of the peculiar velocity than in the case of a large positive
peculiar velocity~\citep{2003JCAP...05..007A}. We ran a sample of different
parameters ($T_e = 2-10$ keV, $v_\gas = -1000 - +1000$ km/sec, $y=10^{-6} -
10^{-4}$), and we find both very little parameter degeneracy with
$\tau_{\CR}$, and that the error-bar on $\tau_{\CR}$ remains $2\cdot 10^{-5}$ within
a factor of 2 for all the clusters in this sample. This also implies that
systematic shifts of the other cluster paramters due to non-removable
contamination should not strongly alter our conclusions.

\begin{figure}
\begin{center}
\includegraphics[width=0.5\textwidth]{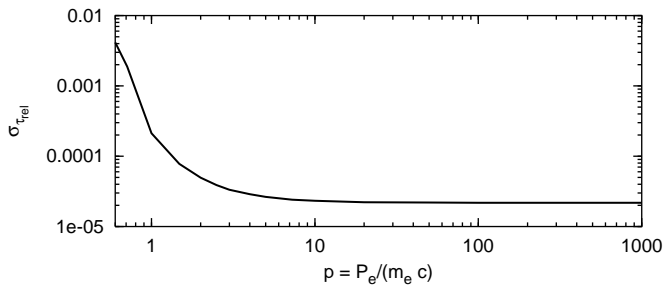}
\end{center}
\caption{The error-bar on $\tau_{\CR}$ as a function of the dimension
less electron momentum, $p$. The error-bar on $\tau_{\CR}$ is independent
of $p$ as long as $p>3$.}
\label{fig:p.dtau}
\end{figure}

As can be clearly seen on figure~\ref{fig:p.dtau} the determination of
$\tau_\CR$ gives much larger error-bars for $p<3$ (as explained in
section~\ref{sec:spectra}), compared to the case $p>3$ with small and constant
error-bars. $\tau_{\CR}$ is therefore practically non-determinable for
$p<1$. However, the unproblematic $p>3$ case is exactly the range we are
interested in.

We therefore conclude that the error-bar on the determination of
$\tau_{\CR}$ will be about $2\cdot 10^{-5}$ for any single typical clusters detectable by
PLANCK, roughly independent of other cluster parameters.

\section{Conclusion}

We estimate the sensitivity of PLANCK to detect the optical depth $\tau_\CR$
of relativistic electrons to be $\sigma_{\tau_\CR} \sim 2\cdot 10^{-5}$ for a
single cluster. This is too insensitive to detect even optimistic senarios for
the relativistic electron population, which give values of $\tau_\CR \sim
10^{-7}$ to $10^{-6}$.

However, there is no significant parameter degeneracy with other cluster
parameters for the measurement of the optical depth $\tau_\CR$ of relativistic
electrons. This allows one to stack the weak signals of the large number of
galaxy clusters to be detected by PLANCK with the thermal SZ effect. Since
PLANCK is expected to observe of the order of $N_{\rm cl} \approx 4\cdot 10^4$
individual clusters \citep[e.g.][]{2002A&A...388..732B}, the expected
sensitivity for the statistical detection of the relativistic SZ signal should
be of the order $\sigma_{\tau_{\CR}}^{\rm stat} = \sigma_{\tau_\CR} \,N_{\rm
cl}^{-1/2} \sim 10^{-7}$. This is sufficient to confirm or refute optimistic
scenarios.

Future experiments with increased sensitivities may be able to test
for even smaller optical depth of relativistic electron populations.
An experiment with 4 observing frequencies at 90, 150, 217 and 270
GHz, and with South-Pole-Telescope\footnote{\tt
http://astro.uchicago.edu/spt/}-like sensitivity of $0.1 \,\mu$K would
allow to probe relativsitic electrons down to $\sigma_{\tau_{\CR}}
\sim 2\cdot 10^{-6}$ for an individual cluster.  Thus optimistic
theoretical scenarios could be tested using a few individual clusters,
while statistical measurements would probe even deep into conservative
scenarios of relativistic electron populations in galaxy clusters.

We conclude that sensitive CMB experiments like PLANCK allow to
constrain or examine the presently poorly known low energy part of the
existing relativistic electron populations in galaxy clusters. Therefore
they provide a valuable vehicle to explore this {\it terra incognita}
in electron energies in between MeV and GeV.

\begin{acknowledgements}
It is a pleasure to thank Nabila Aghanim for initiating this project
and Gerhard B{\"o}rner and Christoph Pfrommer for comments on the
manuscript.  SHH thanks the Tomalla foundation for financial
support. This research was done in the framework of the European
Comunity {\it Cosmic Microwave Background Network}.
\end{acknowledgements}

\bibliography{aamnem99,tae}

\begin{thebibliography}{}

\bibitem[\protect\astroncite{{Aghanim} et~al.}{2003}]{2003JCAP...05..007A}
{Aghanim}, N., {Hansen}, S.~H., {Pastor}, S., {Semikoz}, D.~V., 2003,
\newblock {J. of Cosmology and Astro-Particle Physics} {5}, 7

\bibitem[\protect\astroncite{{Bartelmann}}{2001}]{2001A&A...370..754B}
{Bartelmann}, M., 2001,
\newblock {\aap} {370}, 754

\bibitem[\protect\astroncite{{Bartelmann} \&
  {White}}{2002}]{2002A&A...388..732B}
{Bartelmann}, M., {White}, S.~D.~M., 2002,
\newblock {\aap} {388}, 732

\bibitem[\protect\astroncite{{Blasi}}{2000}]{2000ApJ...532L...9B}
{Blasi}, P., 2000,
\newblock {\apjl} {532}, L9

\bibitem[\protect\astroncite{{Blasi} et~al.}{2000}]{2000ApJ...535L..71B}
{Blasi}, P., {Olinto}, A.~V., {Stebbins}, A., 2000,
\newblock {\apjl} {535}, L71

\bibitem[\protect\astroncite{{Brunetti}}{2002}]{astro-ph/0208074}
{Brunetti}, G., 2002,
\newblock in S. {Bowyer} \& C.-Y. {Hwang} (eds.), {Matter and Energy in
  Clusters of Galaxies}, ASP Conference Series,
\newblock astro-ph/0208074

\bibitem[\protect\astroncite{{Colafrancesco}
  et~al.}{2003}]{2003A&A...397...27C}
{Colafrancesco}, S., {Marchegiani}, P., {Palladino}, E., 2003,
\newblock {\aap} {397}, 27

\bibitem[\protect\astroncite{{Diego} et~al.}{2003}]{2003MNRAS.338..796D}
{Diego}, J.~M., {Hansen}, S.~H., {Silk}, J., 2003,
\newblock {\mnras} {338}, 796

\bibitem[\protect\astroncite{{Dolgov} et~al.}{2001}]{2001ApJ...554...74D}
{Dolgov}, A.~D., {Hansen}, S.~H., {Pastor}, S., {Semikoz}, D.~V., 2001,
\newblock {\apj} {554}, 74

\bibitem[\protect\astroncite{{En{\ss}lin}}{1999a}]{Ringberg99}
{En{\ss}lin}, T.~A., 1999a,
\newblock in P.~S. H.~{B{\"o}hringer}, L.~{Feretti} (ed.), {Ringberg Workshop
  on `Diffuse Thermal and Relativistic Plasma in Galaxy Clusters'}, Vol. 271 of
  {MPE Report}, p. 275,
\newblock astro-ph/9906212

\bibitem[\protect\astroncite{{En{\ss}lin}}{1999b}]{Pune99}
{En{\ss}lin}, T.~A., 1999b,
\newblock in {IAU Symp. 199: `The Universe at Low Radio Frequencies'},
\newblock astro-ph/0001433

\bibitem[\protect\astroncite{{En{\ss}lin} \&
  {Biermann}}{1998}]{1998AA...330...90E}
{En{\ss}lin}, T.~A., {Biermann}, P.~L., 1998,
\newblock {\aap} {330}, 90

\bibitem[\protect\astroncite{{En{\ss}lin} \&
  {Kaiser}}{2000}]{2000A&A...360..417E}
{En{\ss}lin}, T.~A., {Kaiser}, C.~R., 2000,
\newblock {\aap} {360}, 417

\bibitem[\protect\astroncite{{En{\ss}lin} et~al.}{1999}]{1999AA...344..409E}
{En{\ss}lin}, T.~A., {Lieu}, R., {Biermann}, P.~L., 1999,
\newblock {\aap} {344}, 409

\bibitem[\protect\astroncite{{En{\ss}lin} \&
  {Sunyaev}}{2002}]{2002A&A...383..423E}
{En{\ss}lin}, T.~A., {Sunyaev}, R.~A., 2002,
\newblock {\aap} {383}, 423

\bibitem[\protect\astroncite{{Fusco-Femiano}
  et~al.}{1999}]{1999ApJ...513L..21F}
{Fusco-Femiano}, R. {\it et al.},
1999,
\newblock {\apjl} {513}, L21

\bibitem[\protect\astroncite{{Fusco-Femiano} et~al.}{2004}]{astro-ph/0312625}
{Fusco-Femiano}, R. {\it et al.},
2004,
\newblock {\apjl} in press,
\newblock astro-ph/0312625

\bibitem[\protect\astroncite{{Giovannini} et~al.}{1999}]{Giovannini.Pune99}
{Giovannini}, G., {Feretti}, L., {Govoni}, F., 1999,
\newblock in {IAU Symp. 199: `The Universe at Low Radio Frequencies'},
\newblock astro-ph/0006380

\bibitem[\protect\astroncite{{Haehnelt} \&
  {Tegmark}}{1996}]{1996MNRAS.279..545H}
{Haehnelt}, M.~G., {Tegmark}, M., 1996,
\newblock {\mnras} {279}, 545

\bibitem[\protect\astroncite{{Hansen}}{2003}]{astro-ph/0310149}
{Hansen}, S.~H., 2003,
\newblock {New Astronomy} in press,
\newblock astro-ph/0310149

\bibitem[\protect\astroncite{{Hansen} et~al.}{2002}]{2002ApJ...573L..69H}
{Hansen}, S.~H., {Pastor}, S., {Semikoz}, D.~V., 2002,
\newblock {\apjl} {573}, L69

\bibitem[\protect\astroncite{{Itoh} \& {Nozawa}}{2003}]{2003astro.ph..7519I}
{Itoh}, N., {Nozawa}, S., 2003,
\newblock {ApJ Suppl. submitted},
\newblock astro-ph/0307519

\bibitem[\protect\astroncite{{Liang} et~al.}{2002}]{2002MNRAS.337..567L}
{Liang}, H., {Dogiel}, V.~A., {Birkinshaw}, M., 2002,
\newblock {\mnras} {337}, 567

\bibitem[\protect\astroncite{{Petrosian}}{2001}]{2001ApJ...557..560P}
{Petrosian}, V.~., 2001,
\newblock {\apj} {557}, 560

\bibitem[\protect\astroncite{{Rephaeli}}{1995}]{1995ApJ...445...33R}
{Rephaeli}, Y., 1995,
\newblock {\apj} {445}, 33

\bibitem[\protect\astroncite{{Rossetti} \& {Molendi}}{2004}]{astro-ph/0312447}
{Rossetti}, M., {Molendi}, S., 2004,
\newblock {\aap} in press,
\newblock astro-ph/0312447

\bibitem[\protect\astroncite{{Sarazin}}{1999}]{1999ApJ...520..529S}
{Sarazin}, C.~L., 1999,
\newblock {\apj} {520}, 529

\bibitem[\protect\astroncite{{Sunyaev} \&
  {Zeldovich}}{1972}]{1972ComAp...4..173S}
{Sunyaev}, R.~A., {Zeldovich}, Y.~B., 1972,
\newblock {Comments on Astrophysics} {4}, 173

\bibitem[\protect\astroncite{{Wright}}{1979}]{1979ApJ...232..348W}
{Wright}, E.~L., 1979,
\newblock {\apj} {232}, 348

\end{thebibliography}
\bibliographystyle{aabib99}

\end{document}